# Band-Pass and OH-Suppression Filters for the E-ELT – Design and Prototyping

St. Günster* [a], D. Ristau [a], R. Davies [b]
[a] Laser Zentrum Hannover e.V. , Hollerithallee 8, 30419 Hannover, Germany
[b] Max-Planck-Institut für extraterrestrische Physik, 85740 Garching, Germany

## ABSTRACT

Optical filters are used for a variety of purposes at astronomical telescopes. In the near infrared region, from 0.8 to 2.5 µm, bandpass and edge filters are used to separate the different astronomical channels, such as the J, H, and K bands. However, in the same wavelength range light emission generated in the earth's atmosphere is superimposed on the stellar radiation. Therefore, ground based astronomical instruments measure, in addition to the stellar light, also unwanted contributions from the earth's atmosphere. The characteristic lines of this OH emission are extremely narrow and distributed over the complete NIR spectral range.
The sensitivity of future telescopes, like the European Extreme Large Telescope (E-ELT) which is currently being designed by ESO, can be dramatically improved if the atmospheric emission lines are effectively suppressed while the stellar radiation is efficiently transferred to the detector systems. For this task, new types of optical filters have to be developed. In this framework new design concepts and algorithms must be used, combining the measurement needs with practical restrictions. Certainly, the selected deposition process plays the key role in the manufacturing process. Precise and highly stable deposition systems are necessary to realise such filter systems with an appropriate homogeneity. Moreover, the production control techniques must be adapted to match the high level of precision required in the NIR range. Finally, the characterisation set-ups for such filters systems have to be provided. The manufacturing of such a filter system for a feasibility study of an E-ELT instrument is presented. The design development, the deposition with adapted Ion Beam Sputtering deposition plants, and the characterisation of such filters in the J-Band is described.

## 1. INTRODUCTION

When using astronomical telescopes, the near-infrared spectral range from 0.8 to 2.5µm encompasses a suite of important continuum and emission line features from astrophysical objects. These include thermal continuum from hot dust grains and stellar radiation; and at specific wavelengths, absorption features produced in stellar atmospheres, as well as emission lines from molecular and ionised gas in the interstellar medium. Furthermore, there are two particular advantages of observing in the infrared. The first is that the radiation is relatively insensitive to absorption and scattering by interstellar dust grains, so that it can probe into obscured regions that cannot be seen at optical wavelengths (e.g. young star forming regions, or the centres of galaxies). The second, specifically with respect to ground-based telescopes, is that adaptive optics can correct the turbulence induced by the atmosphere and hence recover the full diffraction limit of the telescope, enhancing both resolution and sensitivity.

Astronomical instrumentation is able to transfer and detect near-infrared radiation rather efficiently: the throughput of an imaging camera, including the detective efficiency, can exceed 60%. And because such instruments are cryogenic, operating at 80-100K, they add very little additional thermal background. The main disadvantage of the near-infrared spectral range is the impact of the earth's atmosphere. There is a high transmission only for specific bands (e.g. the J-band at roughly 1.15-1.35µm, the H-band at 1.45-1.8µm, and the K-band at 1.95-2.5µm). In addition, OH radicals created by reactions between ozone and hydrogen high in the atmosphere produce so-called airglow emission, which is manifested as numerous strong emission lines throughout the near-infrared spectral range. A typical atmospheric transmittance spectrum together with the disturbing OH emission lines is shown in Figure 1. At modest spectral resolution ($\lambda/\delta\lambda \sim 3000$), the intensity of the strongest OH lines is more than 1000 times that of the background level between them. Because their intensity varies on timescales of minutes, even for spectroscopy special techniques are sometimes required to subtract them out fully [1]. For near-infrared astronomical imaging, the OH lines have a particularly strong impact, severely limiting the sensitivity within the passband. A future task should therefore be the efficient suppression of the OH emission lines in the earth's sky (Figure 1). A smart suppression of OH lines together with an optimised transmittance band will increase the sensitivity of ground-based instruments by a significant factor, leading to a valuable improvement in observing efficiency (for example, reducing the background by a third would enable one to reach the same sensitivity in half the observing time). The progress in optical deposition technology could be used to contribute with solutions for the given task [2-4].

* E-Mail: sg@lzh.de ; Tel : ++ 49 511 2788 444 ; Fax : ++ 49 511 2788 100

For the upcoming European Extremely Large Telescope E- ELT [5], which is currently being designed by ESO, the sensitivity of the imaging camera MICADO [6] can be dramatically improved if the atmospheric emission lines are effectively suppressed while the radiation of interest is efficiently transferred to the detector systems. For this task, new types of optical filters have to be developed.

This study summarises results of a feasibility study of coating deposition and coating design development for a typical band pass filter and the first step in an OH suppression filter system. Design solutions for the band pass and more complex OH suppression filters were developed by Monte-Carlo simulation methods and needle algorithms.

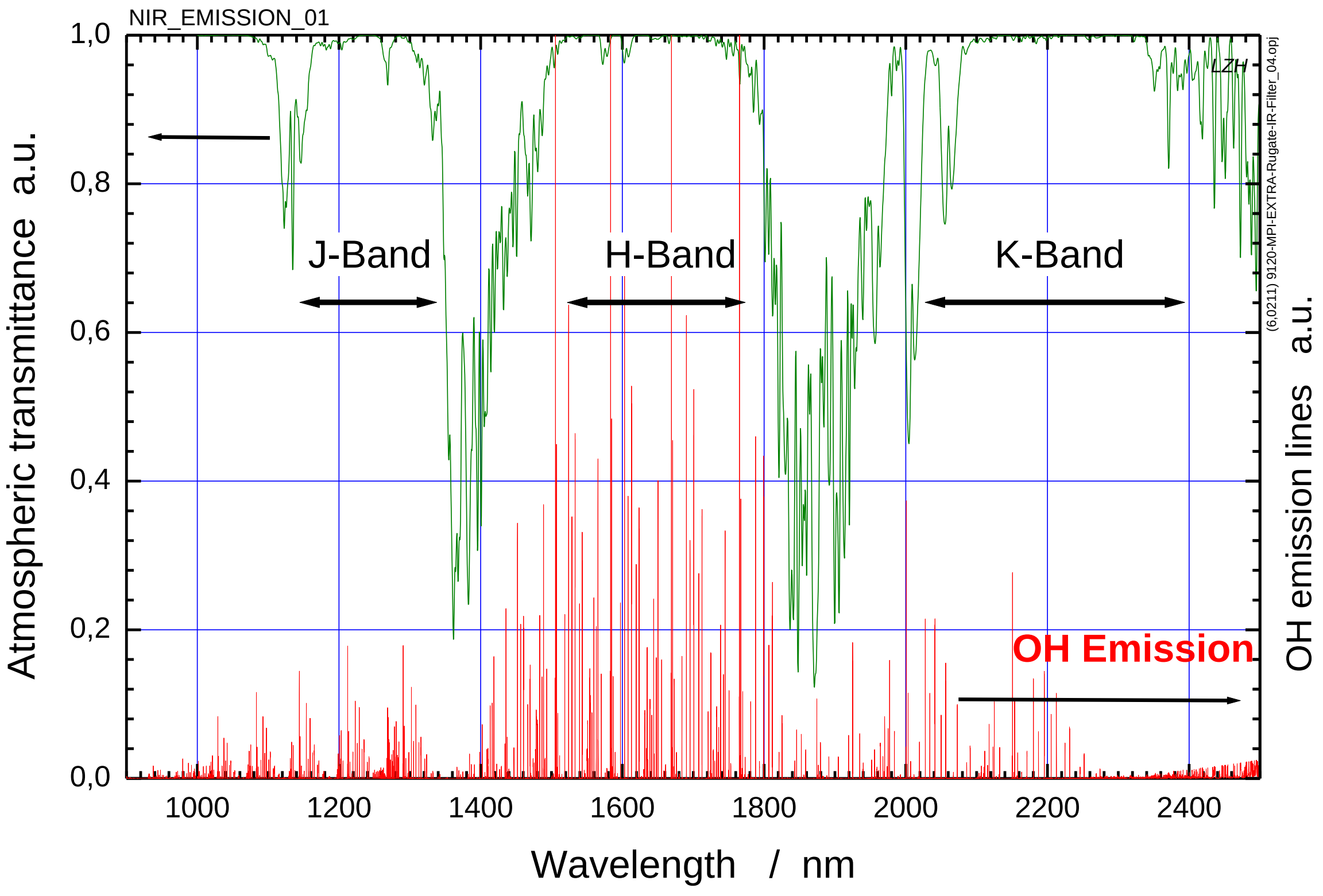

Figure 1: Optical transfer function of the earth's atmosphere (green line) and OH emission lines (red lines, arbitrary scaling) in the NIR spectral range. The standard astronomical detection bands J, H, K are indicated.

## 2. PROBLEM DEFINITION

All astronomical instruments include a set of spectral filters that provide a range of different band passes. The coatings of the filters must provide spectral ranges with a sufficiently high blocking value outside the bandpass combined with high transmittance within the specified regions. Special systems are OH suppression. They should block the molecular emission generated in earth's atmosphere. In addition to these optical properties, the tolerable wavefront errors are low and stress effects onto the substrate caused by the optical coatings must be minimised. In the presented study, three questions were investigated:

1. *Band Pass Filter for the J-Band*: Can J-Band filters be produced safely, and are they stable at liquid $N_2$ temperatures? A coating design was developed and the coating was produced on test substrates. The coating was characterised with respect to the optical function, spatial homogeneity and mechanical stress. Since most of the components will be operated at liquid $N_2$ temperature, the optical behaviour at -196°C was measured in a cryostat.
2. *OH Suppression filter for the J Band / Narrow Band Passes*: Can narrow band blocking structures in the J Band be safely and accurately realised? In a first approach designs were developed to suppress in the J Band region the most severe OH emission lines with narrow band pass filter. The coatings were deposited and characterised.
3. *OH Suppression filter for the J Band / Complex Targets*: Are highly complex design solution for OH suppression feasible? The increase of sensitivity of astronomical instruments can be realised by suppression of the OH lines. An intelligent combination of suppression and transmittance in the J Band region was investigated. New designs were developed under consideration of both a complex suppression scheme and the production possibilities of advanced deposition systems. These design solutions could be the basis for further instrumental improvement.

## 3. EXPERIMENTAL IMPLEMENTATION

### 3.1 Deposition techniques

The targeted coating systems can only be realised with a structure of dense layers. Thermal shift and contamination must be excluded or at least minimised. Therefore, only IAD or Sputtering deposition processes can be used for the coating manufacturing. Furthermore, it is expected that the optical specifications call for narrow band structures. Thus rugate filter technology and non quarterwave deposition will be important for the realisation of such coatings. Highly advanced rugate filter technology was developed and established by the Laser Zentrum Hannover. For such systems Ion Beam Sputtering (IBS) technology was used. Utilising a smart combination of ion source and target material movement, discrete material systems, mixed materials, as well as gradient index coatings can be deposited routinely. The LZH systems are equipped with sophisticated optical online monitoring tools, which are employed for automatic process control. In combination with simulation and optimisation tools, coating designs can be investigated with respect to their production yield prior to the deposition [7]. The BBM online monitoring system allows the precise deposition of nearly any thicknesses unless some spectral change is generated in the online spectra. At this stage, no special measures have been taken with respect to coating homogeneity, which will be of importance for the production of larger optics sizes (e.g. 10cm rather than 2.5cm diameter).

### 3.1 Measurement techniques

Witness samples were measured by commercial spectrometers (Perkin Elmer λ900 / λ1050). Special accessories were used to measure the optical performance at low temperature ($N_2$ bath cryostat, LZH) and to control spatial coating performance on the substrate surface (AHM set-up, LZH). In addition, the thermal shift was quantified in the temperature range between 20 and 150 °C. Components with steep edges and narrow structures were deposited. For the precise measurement of such components, standard commercial spectrophotometers must be improved with respect to their beam quality (divergence spatial precision) for future applications. In addition, defect investigation of the coating system was performed with dark field microscopy and scattering measurements. For selected coatings system electron micrographs were taken. The mechanical stress of the samples was measured with interferometric methods (MW100, Möller Wedel).

### 3.1 Design optimisation tools

The creation of multilayer designs is a highly complex task. Computer based algorithms were utilised in this study. Designs were developed and optimised with the LZH software package SPEKTRUM [8] using mainly Monte Carlo optimisation methods. Especially for the development of rugate filters the SPEKTRUM algorithm provides useful design tools. Furthermore, the OPTILAYER [9, 10] software was employed for solutions generated with the needle algorithm. The needle algorithm generates multilayer structures corresponding to a given target value by introducing new layers into the optimised multilayer system automatically. This new layer – the needles – are placed at the most sensitive sites in the multilayer structure.

## 4. RESULTS

### 4.1 Coating J-Band Filter

J-band filter should transmit efficiently in the J-Band (app. 1147- 1420nm) and block the rest of the NIR and visible range effectively (Figure 2). The wavelengths over which the blocking must be performed extends across 0.7-2.5µm, a range that is defined by the specific requirements of the instrument and detectors. For MICADO, at the short wavelength end other optical components in the telescope and instrument, such as a dichroic to transfer the optical radiation to an adaptive optics wavefront sensor, provide sufficient blocking. The limit at the long wavelength end is set by the cut-off in the detector sensitivity. A design for the J-Band filter was developed on the basis of discrete multilayers. The system has an overall physical thickness of app. 20 µm. The realised component is in very good agreement with the coating design (Figure 3, Figure 4). A blocking better than OD3 was planned and realised in the block band of the J-Band filter (Figure 3). Repeated thermal cycling in the $N_2$ bath does not damage the sample. The coating is hard and stable. The thermal shift between spectra measured at room temperature and liquid nitrogen temperature was found to be 2 – 3 nm (Figure 5). Microstructure investigations with electron micrographs exhibit perfect interfaces without defects and without any increase of surface roughness for increasing layer number (Figure 6).

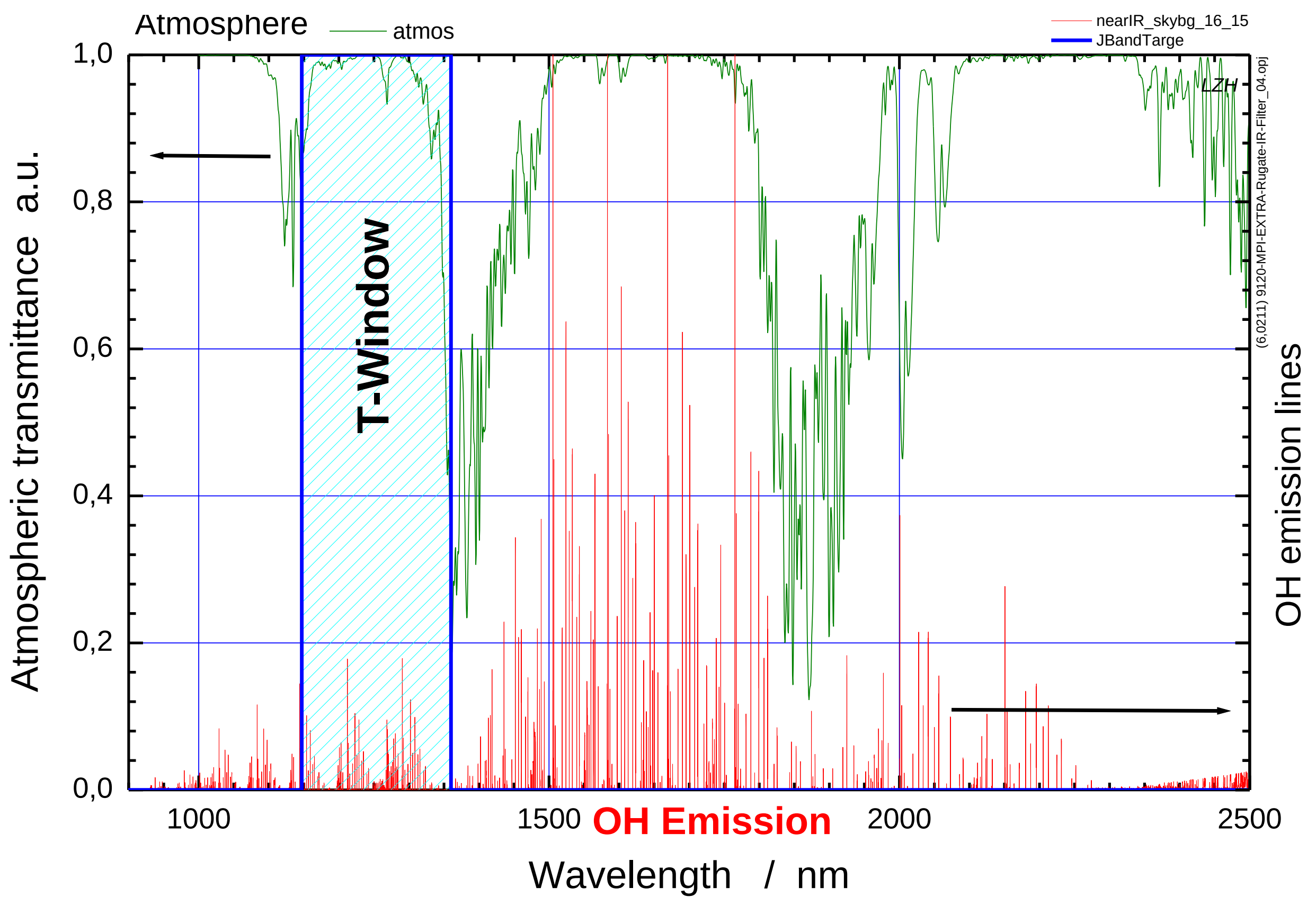

**Figure 2:** Target specification for the J-Band Filter in the NIR spectral range. The green line shows the atmospheric transmission and the red lines the atmospheric OH emission (arbitrary scaling).

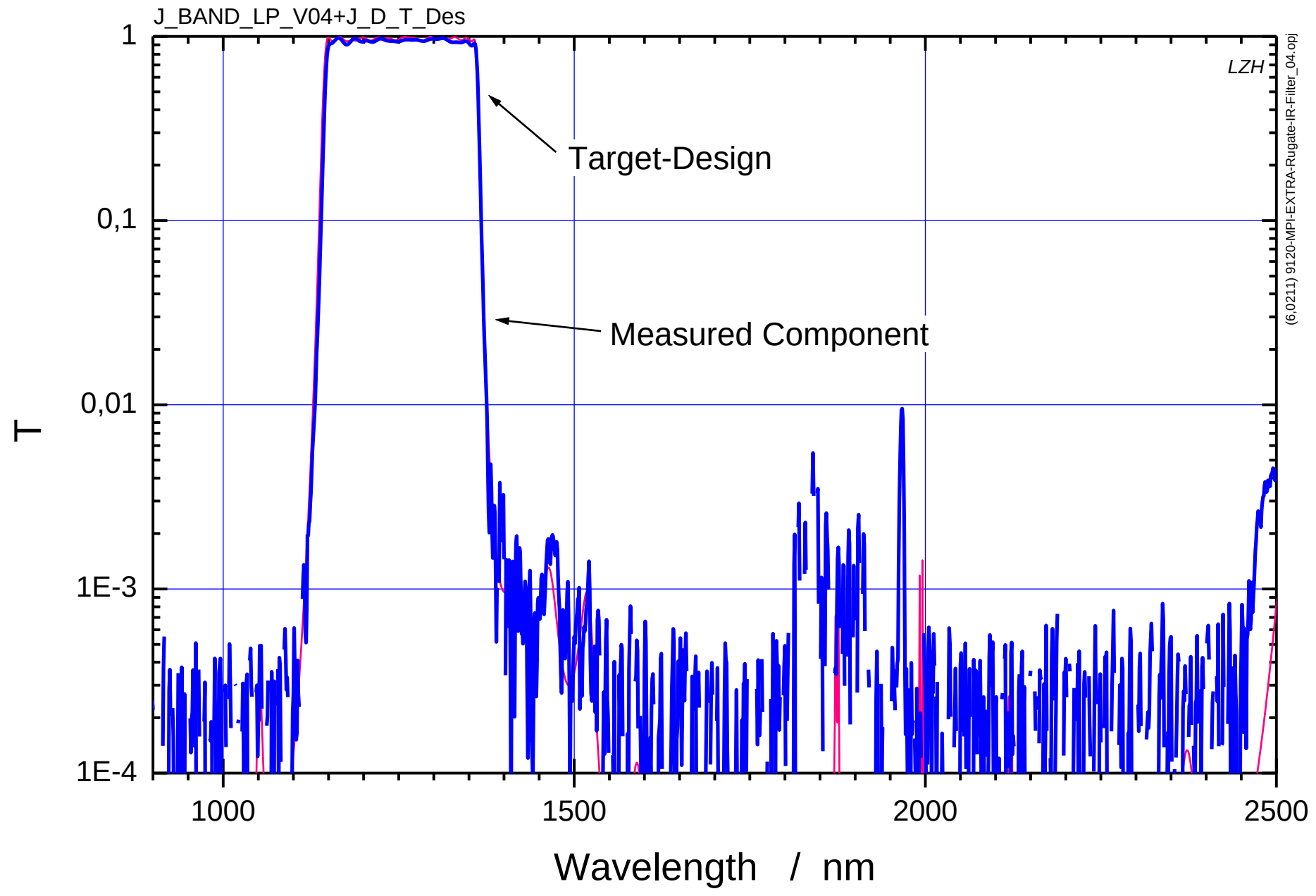

**Figure 3:** Comparison of deposited J Band Filter with the target design. The agreement is excellent. Blocking values better than OD3 are achieved in the outer band regions.

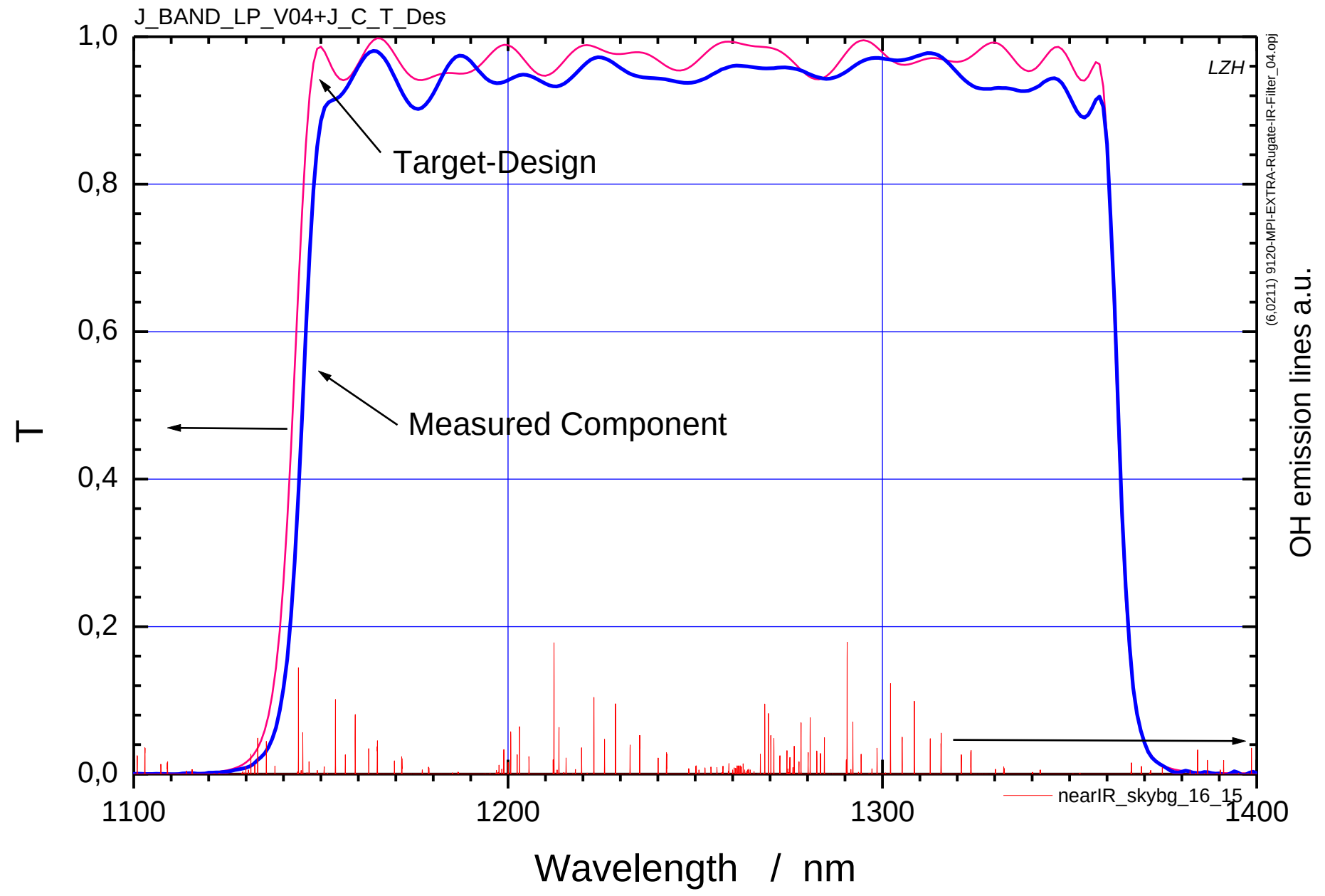

**Figure 4:** Detailed **c**omparison of deposited J Band Filter with the target design in the pass band region. The agreement is excellent.

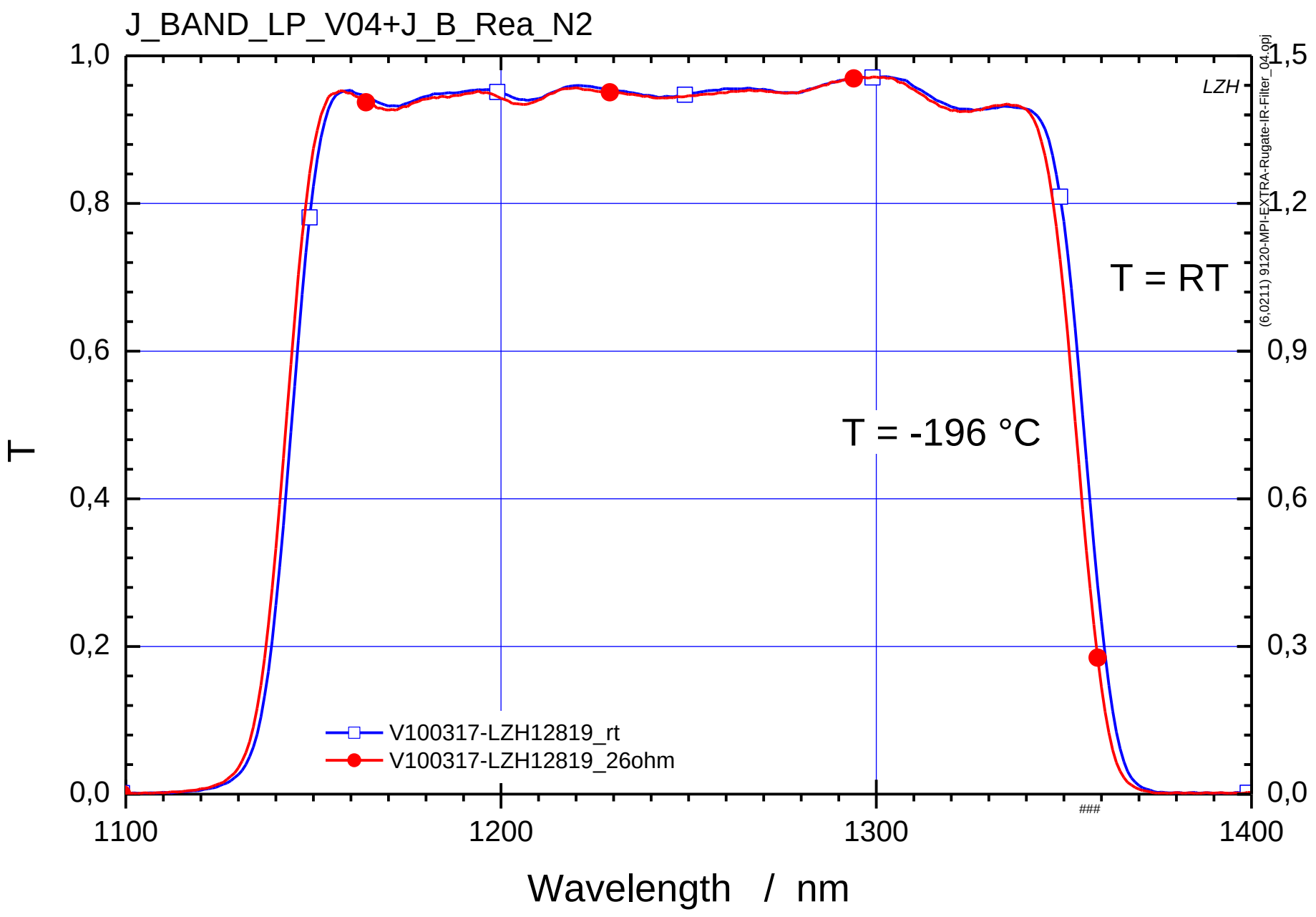

**Figure 5:** Comparison of J Band filter performance in the pass band region for a measurement at room temperature and at liquid nitrogen temperature. The thermal shift has a value of app. – (2-3) nm for the given coating design. The shift is caused by the thermal properties of the materials and fully reversible.

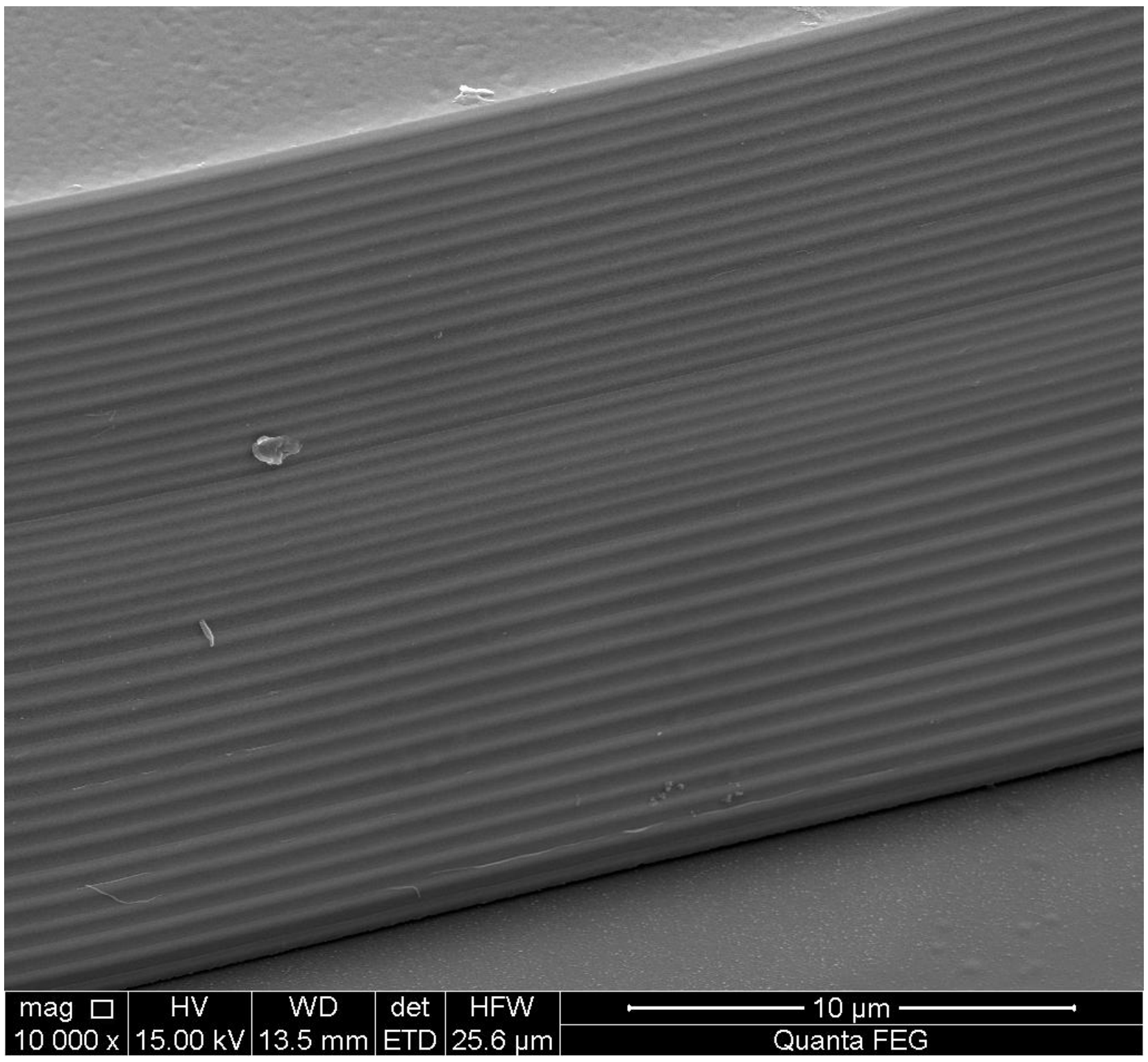

**Figure 6 :** J Band filter structure in an electron micrograph picture. The interfaces are defect free. This is a superior property of the IBS deposition process used.

### 4.2 Coating OH Suppression filter in J-Band Filter

OH emission should be partly suppressed in the J-Band region. If one wishes to keep the filter design relatively simple, estimations for the best instrumental performance yield a three narrow band blocking filter structure (Figure 7). Discrete layer and rugate filter design approaches were investigated. It should be mentioned, that the BBM System allows the deposition of nearly all layer thicknesses – one is not restricted to quarterwave layers etc. The solutions for the given narrow band pass were generated with discrete and rugate designs. Due to the gradient index change in the rugate filters, their total physical thickness is increased with respect to a discrete system. For this reason, the discrete solution with non quarterwave layers was preferred for realisation in this study. The total physical thickness of the design used is about 16 µm. The comparison between the target design curve and the measured performance of the final component is displayed (Figure 8). The agreement between target and measurement is fine. The curve structure is maintained. Some systematic errors manifested by a wavelength shift to shorter wavelengths are due to small errors in the refractive index of the coating material during the deposition process. It is important to bear in mind that online monitoring during deposition is performed in the range from 400 to 950 nm (i.e. there is no overlap with the bandpasses of the filter itself). Features in the NIR range are predicted by calculating methods based on the coating design. Small shifts such as these can be compensated in the instrument itself by mounting the filter at an angle slightly offset from the design specification of 10°. Importantly, repeated thermal cycling in $N_2$ bath does not damage the coating or alter the spectral performance. Again the thermal shift is measured in the range from 0 to 150°C. The measured thermal shift is vanishingly small. The comparison of spectra measured at room temperature and at Liquid $N_2$ temperature exhibit a shift to lower wavelengths of about 2-3 nm for a temperature step of about 200°K. (Figure 10)

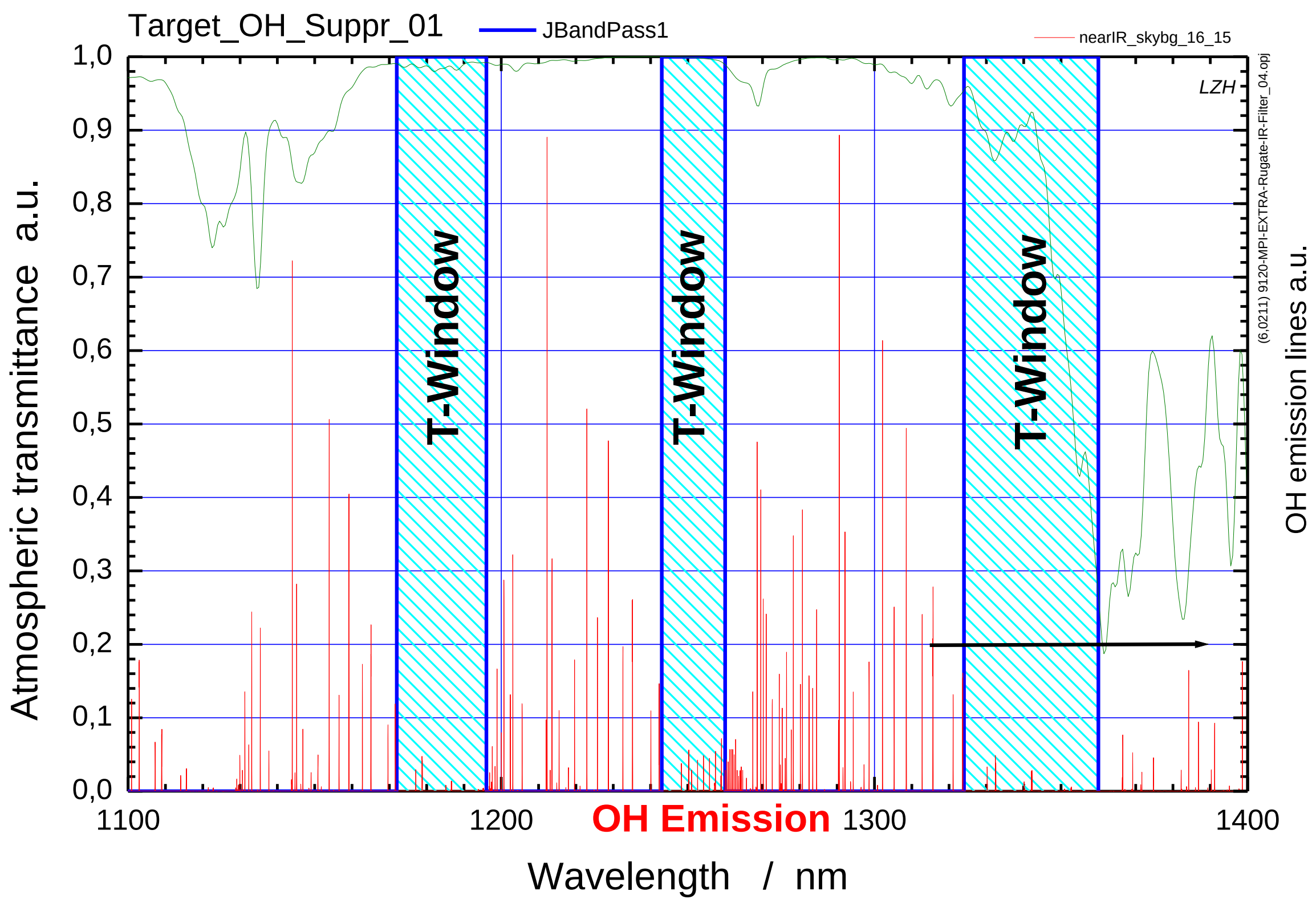

**Figure 7:** Target specification for the OH Suppression V1 Filter in the NIR spectral range: 3 Narrow transmission bands, with blocking bands between them.

* E-Mail: sg@lzh.de ; Tel : ++ 49 511 2788 444 ; Fax : ++ 49 511 2788 100

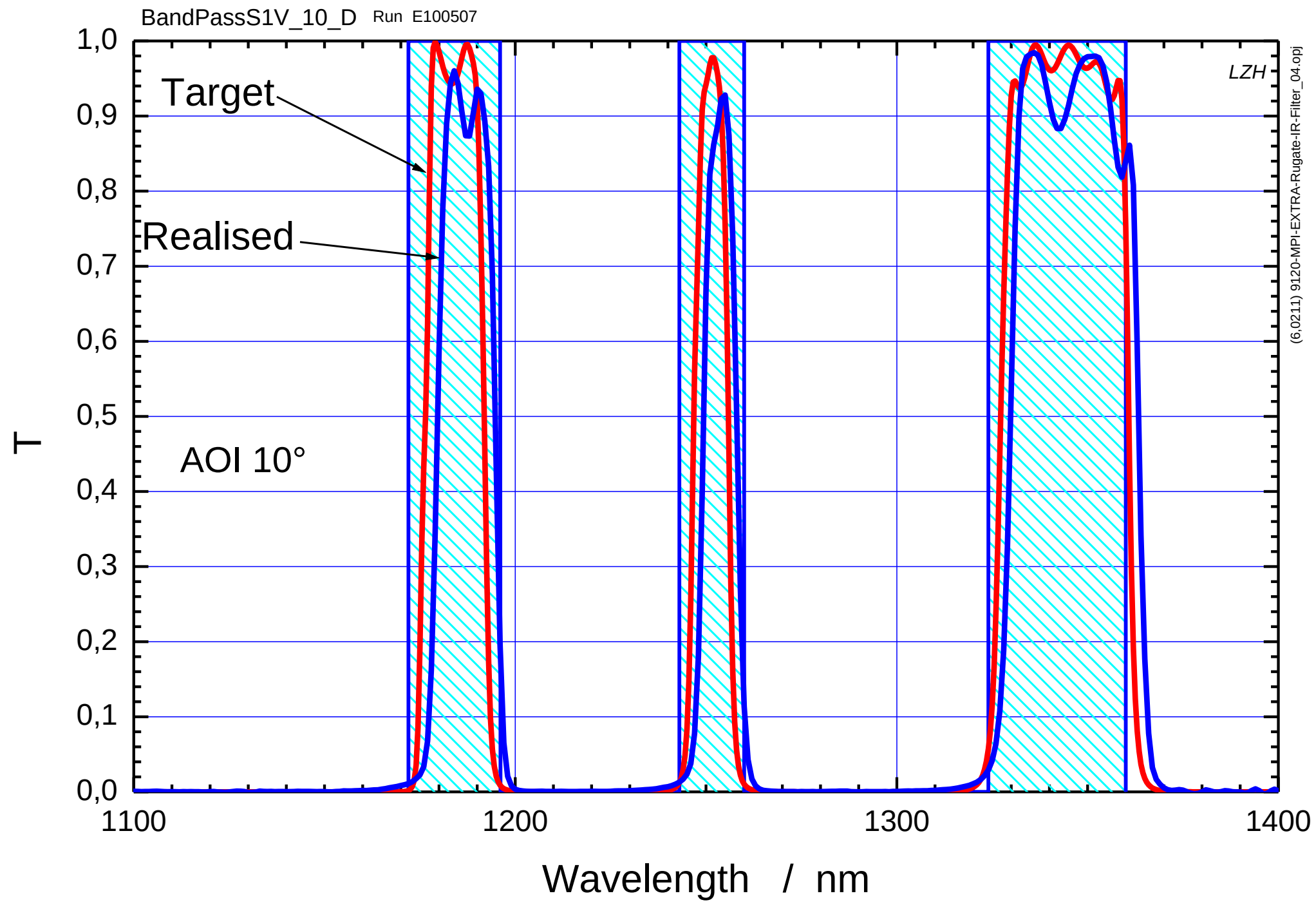

**Figure 8:** Comparison of deposited OH suppression filter. The blocking bands, the target design curve and the measured transmittance of the realized design are displayed.

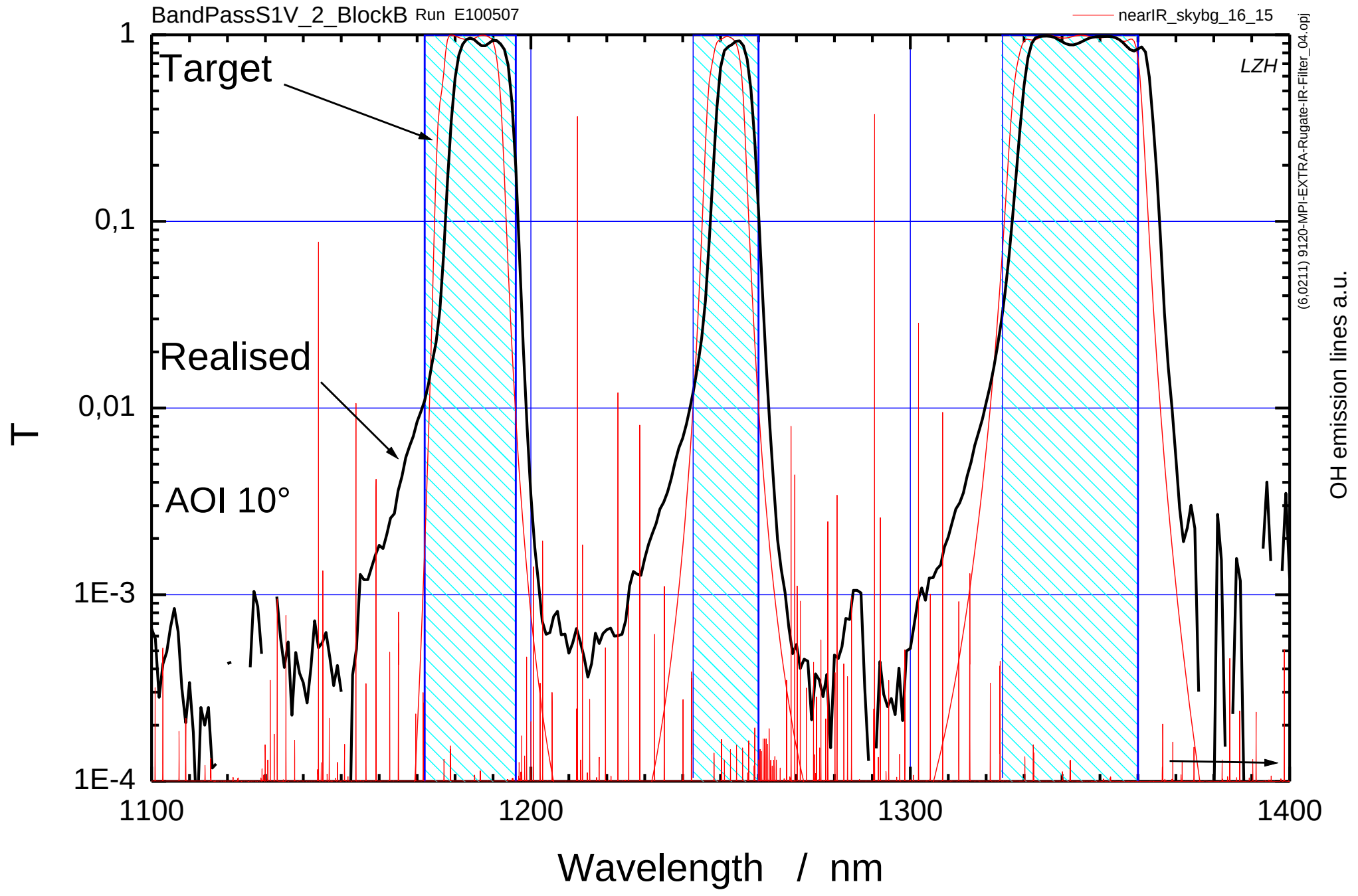

**Figure 9:** Survey of the blocking of OH suppression filter. The blocking bands, the target design curve and the measured transmittance of the realized design are displayed.

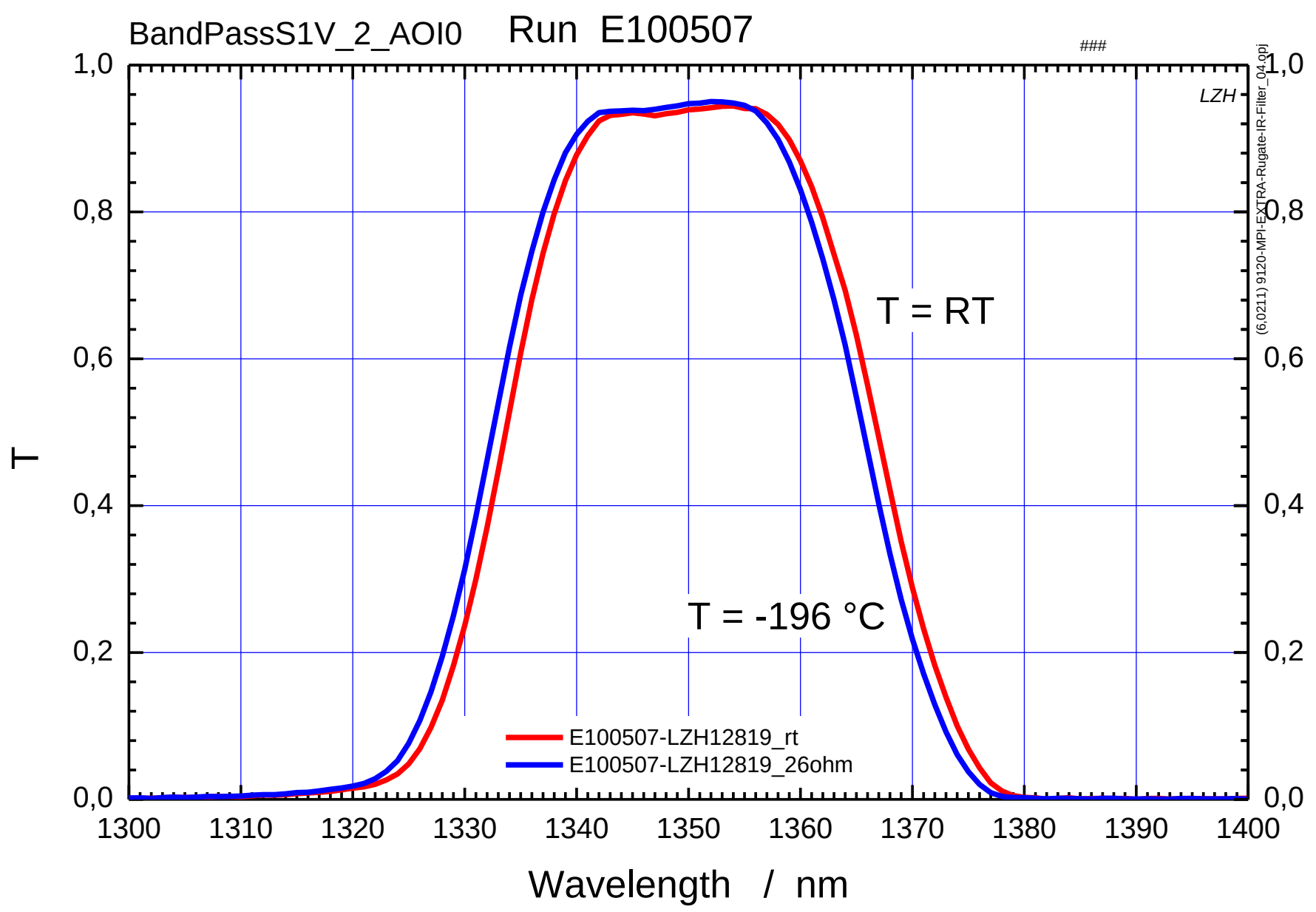

**Figure 10:** Comparison of transmittance spectra at room and liquid $N_2$ temperature of the OH suppression filter. The thermal shift has a value of app. – (2-3) nm for the given coating design. The shift is caused by the thermal properties of the materials and fully reversible.

### 4.3 J Band Suppression filter- Advanced solution

A significant increase in the instrumental performance for a telescope in the NIR range can be realised by an effective suppression of the OH bands. The solution investigated above, and most solutions proposed in the literature, are currently based on narrow band blocking, or the coupling of rugate filter systems etc. In principle, a full blocking of the OH emission lines would be preferable. A wavelength dependent "gain flattening" structure might be an alternative solution. The exact specification must be calculated with respect to detector sensitivity, emission line sensitivity and transfer function for the full instrument. In this framework we have investigated whether a complex suppression filter following an "inverse" structure of the OH emission line can be designed. This means, the filter should show a high transmittance at the wavelength with low OH emission and it should show a higher blocking for wavelength with high OH emission intensity. To solve this problem the OH emission spectrum was inverted. A design will not be possible with the full fine structure of the emission lines. Therefore the emission lines were smoothed to produce a more rough structure (Figure 11). These target numbers were used to develop a design reflecting the wavelength dependent behaviour. A design could be successfully developed. However, the number of small layers inside the layer system is high. In particular, because of the high total physical thickness, a special production scheme for the thin layers must be used for the deposition of such a system (Figure 12).

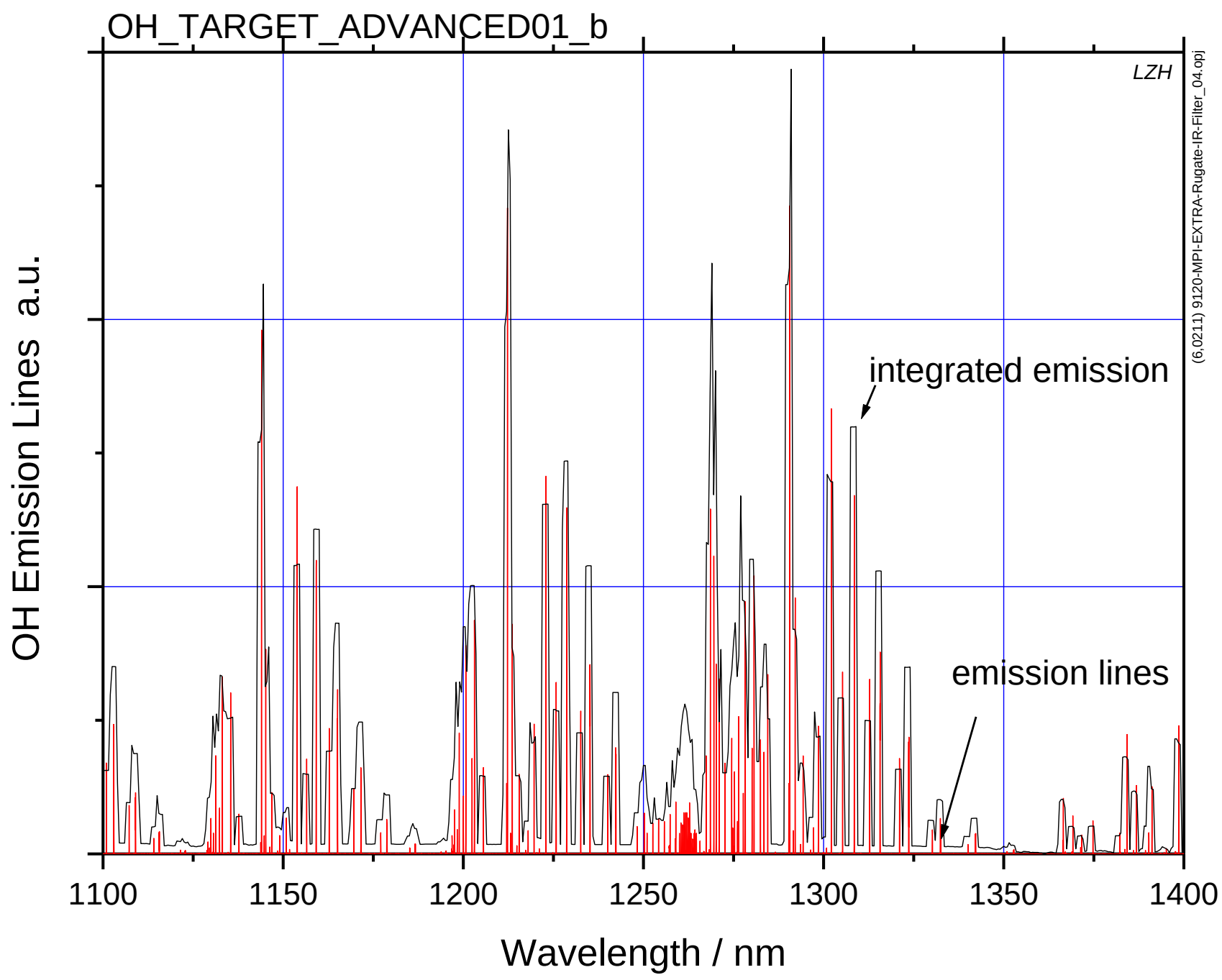

**Figure 11:** OH Emission lines and integrated emission as basis for a complex design development for a wavelength dependent OH emission suppression.

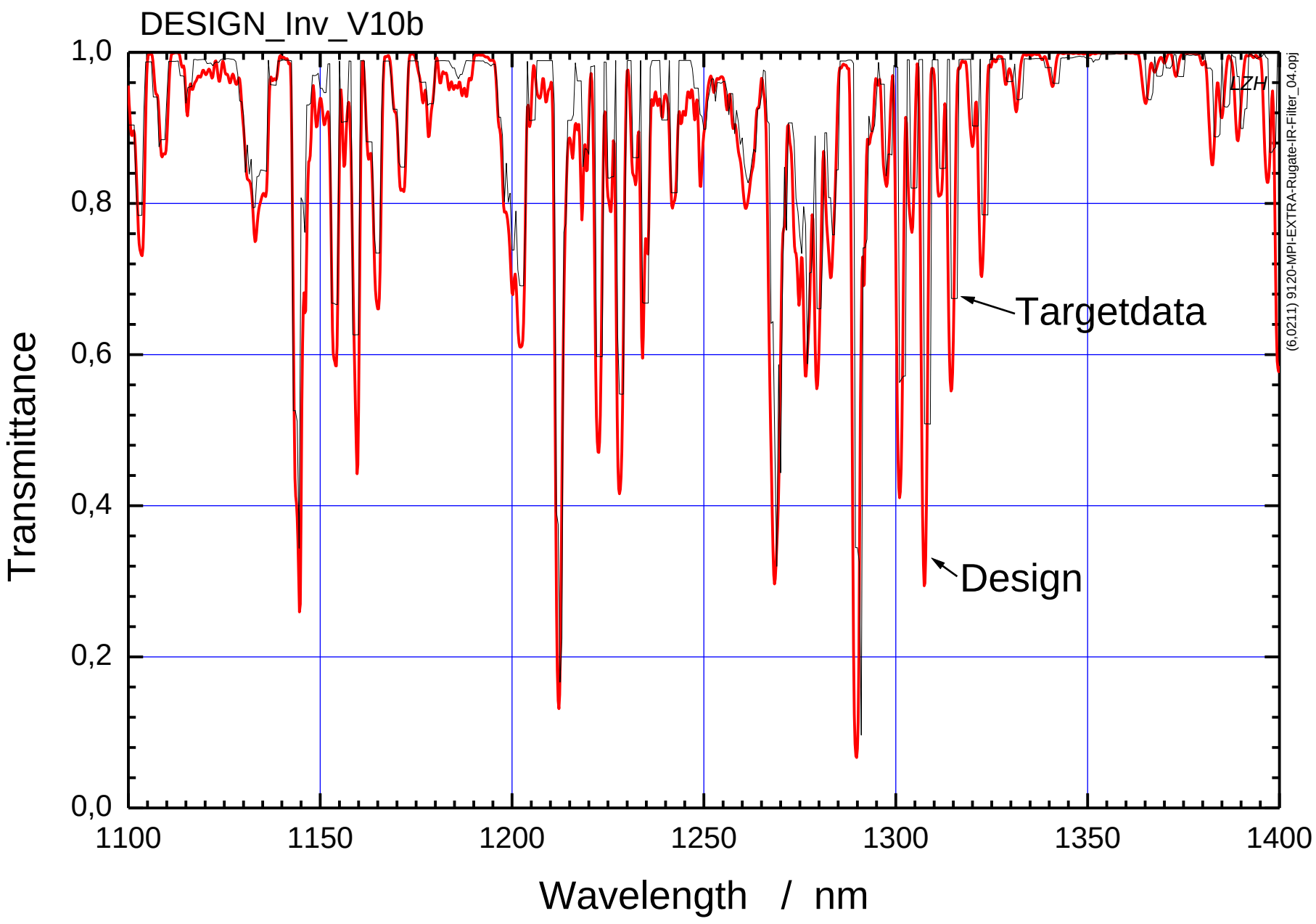

**Figure 12:** Design of a wavelength dependent OH emission suppression system. The complex design has an overall thickness close to 100 µm.

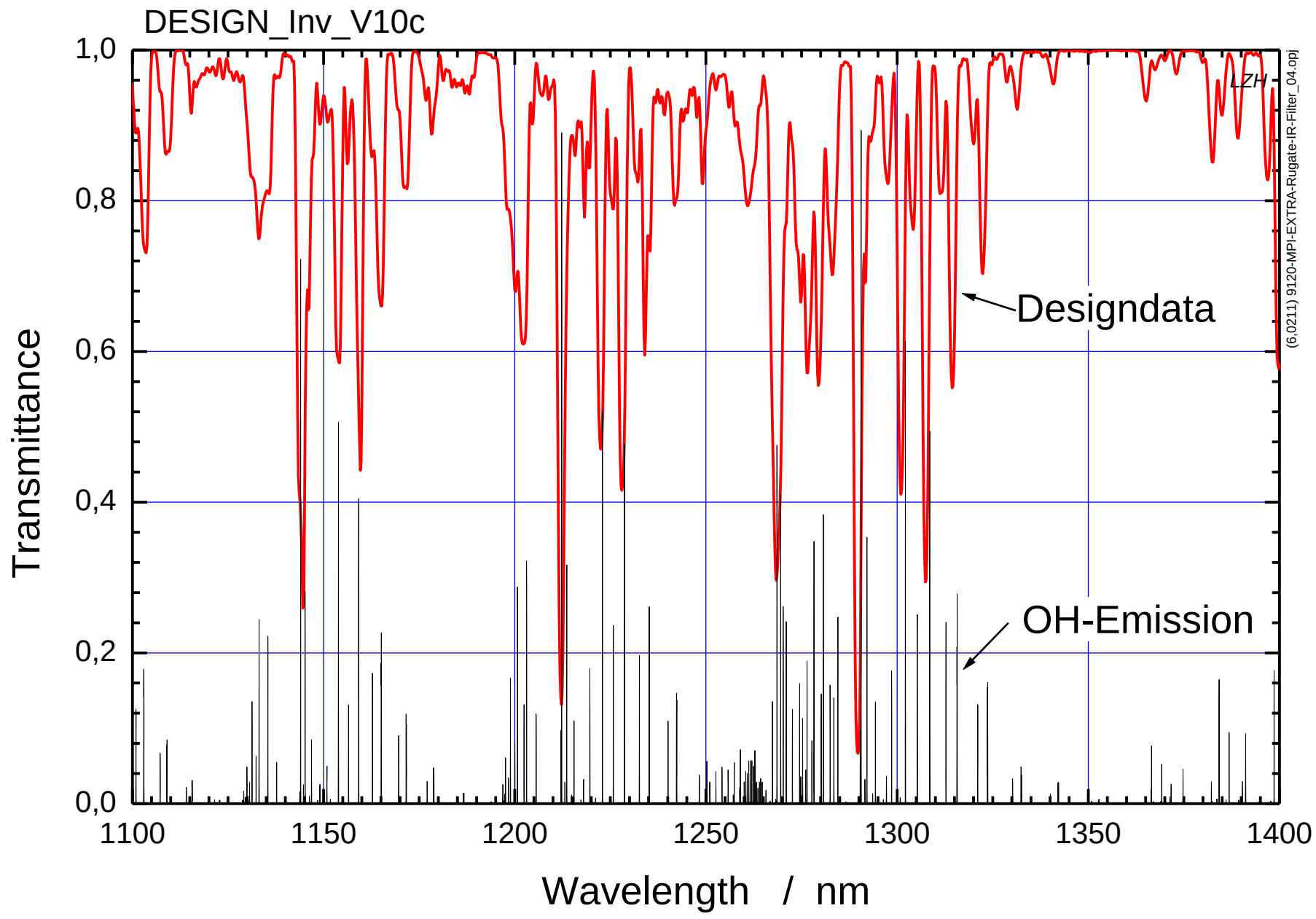

**Figure 13:** Design of a wavelength dependent OH emission suppression system. The complex design has an overall thickness close to 100 µm.

## 5. SUMMARY

A smart suppression of OH emission lines in the NIR spectral range in telescope systems could dramatically improve the performance of ground-based instruments. Progress in deposition technology and deposition process control opens today new optical coating solutions. Online monitoring techniques and the progress in rugate filter technology make new solutions for the next generation of optical coatings for the E-ELT possible. In this study prototype systems of J-Band pass filters and a three narrow band OH blocking filter for the J band were designed and manufactured. The spectral performance was measured and is in excellent agreement with the target values. The stability with respect to thermal cycling down to $N_2$ temperature was measured. The thermal shift is low and could be considered in the optical design process. In addition, highly complex solutions were investigated to create OH suppression filters in an "inverse" transmitting scheme. It was shown that such a design can be formed. The manufacturing of such systems will be challenging due to the high total physical thickness. An individual controlling scheme for the layers will be necessary. From the present results it can be concluded that new and improved coating systems will be available for the E-ELT.

## ACKNOWLEDGEMENTS

The authors gratefully acknowledge the deposition of optical coatings at the Laser Zentrum Hannover by Eugen Buterus and Sonja Wolpers.